\documentclass[12pt]{article}
\usepackage{epsfig}
\usepackage{epsfig,graphicx,psfrag}       

\newcommand{\be}{\begin{equation}}
\newcommand{\ee}{\end{equation}}
\newcommand{\ba}{\begin{eqnarray}}
\newcommand{\ea}{\end{eqnarray}}
\newcommand{\nn}{\nonumber}

\newcommand{\MeV}{{\rm MeV}}
\newcommand{\GeV}{{\rm GeV}}
\newcommand{\ice}[1]{\relax}

\newcommand{\ov}[1]{ \overleftarrow{#1} }
\newcommand{\MSbar}{\overline{\rm MS}}
\def\pslash{\rlap{\hspace{0.02cm}/}{p}}
\begin{document}
\thispagestyle{empty}

\begin{flushright}
DESY 07-038 \\
SI-HEP-2006-17\\
March 2007
\end{flushright}
\phantom{}
\vspace{.2cm}
\begin{center}
{\Large\bf Analyzing $B_s - \bar B_s$ mixing: \\
Non-perturbative contributions to bag parameters from sum rules}\\[1truecm]
\vspace{.25cm}

T.~Mannel$^{a}$, B.D.~Pecjak$^b$, A.A.~Pivovarov$^{a,c}$\\[1mm]
${}^a$Fachbereich 7 (Physik)
Theoretische Physik I
Universit\"at Siegen\\
Emmy Noether Campus
Walter Flex Strasse 3
D-57068 Siegen, Germany 

$^{b}$ Theory Group, Deutsches Elektronen-Synchrotron DESY, 
22603 Hamburg, Germany

${}^c$~Institute for Nuclear Research of the\\
Russian Academy of Sciences, Moscow 117312, Russia

\end{center}

\vspace{1truecm}
\begin{abstract}
We use QCD sum rules to compute matrix elements of the  
$\Delta B=2$ operators appearing in the heavy-quark expansion 
of the width difference of the $B_s$ mass eigenstates. 
Our analysis includes the leading-order operators $Q$ 
and $Q_S$, as well as the subleading operators $R_2$
and $R_3$, which appear at next-to-leading order in 
the $1/m_b$ expansion.
We conclude that the violation of the 
factorization approximation for these matrix
elements due to non-perturbative vacuum condensates is as low as 1-2\%. 
\end{abstract}

\newpage
\tableofcontents

\newpage
\section{Introduction}
\label{sec:Introduction}

The phenomenon of flavour mixing has been intensively investigated
over the
last decades. The standard model of particle physics provides us with a
parameterization of flavour physics which is compatible with all data
taken up to now. 
However, we are still lacking a fundamental theory of flavour, explaining
the three-family structure, the masses and mixings and CP violation.

The phenomenology of flavour mixing has a few peculiarities. In the 
standard model
the only source of  flavour mixing 
originates from the ``mismatch'' between the
two
mass matrices for the up and the down quarks, which is encoded in 
the relative rotation
between the eigenbases 
of these matrices given by the CKM matrix. The mass matrices
are induced by Yukawa couplings 
to the Higgs particle, which hints at a relation between
electroweak symmetry breaking and  the origin of flavour.

CP violation in the standard model is 
related to  an irreducible phase in the CKM
matrix, which can appear 
for at least three 
generations~\cite{Kobayashi:1973fv}. Putting aside the still unsolved
mystery of strong CP violation~\cite{Peccei:1977hh}, this leads 
to a few interesting conclusions which are
confirmed by observation. One of these conclusions 
is the strong suppression of
(CP violating) electric dipole moments of quarks 
and leptons, which is compatible
with data. However, in a generic parameterization 
of ``new physics'' contributions
it is hard to avoid electric dipole moments 
exceeding the experimental limits by
orders of magnitude.

A further peculiarity of the standard parameterization 
of flavour physics is the
suppression of  ``flavour changing neutral currents'' (FCNC's) by 
the GIM mechanism~\cite{Glashow:1970gm}, 
which has its root in the unitarity of 
the CKM matrix. In particular,
FCNC processes with $\Delta B = 2$, $\Delta S = 2$ 
have been intensively investigated,
 while $\Delta C = 2$ processes have not yet been observed, in
accordance with the very strong GIM suppression predicted by the standard
model.

Especially in the systems of neutral $B$ mesons the theoretical description is
simplified by the fact that the mass difference 
in these systems is dominated by
the short distance contribution of the top quark. 
Furthermore, the width difference,
which is expected to be sizable in the $B_s$ system, can be computed in the
heavy-quark expansion~\cite{Manohar:2000dt}.

The width difference $\Delta\Gamma$ between the $B_s$ mass eigenstates
is determined by the off-diagonal
matrix element $\Gamma_{12}$ of 
the $\Delta B=2$ transition operator ${\cal T}$
through $\Delta\Gamma=-2\Gamma_{12}$ where 
\be
\Gamma_{12}=\frac{1}{2M_{B_s}}\langle \bar B_s|{\cal T}|B_s \rangle
\ee
and $M_{B_s}$ is the $B_s$ meson mass.
The $\Delta B=2$ transitions are initiated by a flavour changing
neutral current and occur only at the loop level in the 
standard model.  Therefore the transition operator ${\cal T}$ 
is a complicated, non-local object. The main problem however is 
the treatment of mesons as bound states of 
QCD, which involves dynamics in the infrared strong coupling regime,
where a perturbative treatment is not possible.
In the heavy-quark expansion the off-diagonal
matrix element $\Gamma_{12}$ can be expanded as a series 
in inverse powers of the $b$-quark mass as 
\be\label{eq:transop}
\langle \bar B_s|{\cal T}|B_s \rangle=\sum_n \frac{C_n}{m_b^n}
\langle \bar B_s|{\cal O}_n^{\Delta B=2}|B_s \rangle
\ee
where the Wilson coefficients $C_n$ are calculable in 
perturbation theory~\cite{hagelin:81}. In this formulation
all the non-perturbative physics is contained in the matrix elements
of the local $\Delta B=2$ operators ${\cal O}_n^{\Delta B=2}$.
At leading order in $1/m_b$ the transition operator ${\cal T}$
involves two four-quark operators 
\begin{eqnarray}
Q&=&(\bar b_i s_i)_{V-A}(\bar b_j s_j)_{V-A} \\
Q_S&=&(\bar b_i s_i)_{S-P}(\bar b_j s_j)_{S-P}
\end{eqnarray}
with $i$ a color index. The notation is such that
$(\bar b_i s_i)_{V-A}=\bar b_i \gamma_{\mu}(1-\gamma_5)s_i$
and $(\bar b_i s_i)_{S-P}=\bar b_i (1-\gamma_5)s_i$.
At next-to-leading order in $1/m_b$ the transition operator
involves five new (subleading) operators.
The complete list of subleading operators and different choices of  
basis can be found in~\cite{beneke:96,lenz:06}.  
We shall focus on the operators involving an extra covariant
derivative acting on the strange-quark field, of which there
are four.  Neglecting higher-order terms in the $1/m_b$ expansion
these can further be reduced to the two operators
\begin{eqnarray}
\label{R2op} R_{2}&=&\frac{1}{m_b^2}
(\bar b_i {\ov D_{\mu}} 
{D}^\mu s_i)_{V-A}
(\bar b_i s_i)_{V-A} \\
\label{R3op}
R_{3}&=&\frac{1}{m_b^2}
(\bar b_i {{\ov D}}_\mu
{D}^\mu s_i)_{S-P}
(\bar b_i s_i )_{S-P} 
\end{eqnarray}
with $D_\mu=\partial_\mu-ig_s A_\mu$ the covariant derivative.  
The subleading operators should be understood in 
HQET even though they are written
formally in terms of full QCD fields. This means
that the covariant derivative acting on the $b$-quark field in 
(\ref{R2op})-(\ref{R3op}) can be
replaced by $m_b v$ with $v$ the velocity of the heavy $b$-quark, 
making explicit that the subleading operators 
$R_2$ and $R_3$ are suppressed only by one power of $1/m_b$.

The standard parameterization of the matrix
elements of these operators is obtained
through the vacuum saturation approximation~\cite{Gaillard:1974hs}
with bag parameters $B_i$ controlling the accuracy of the factorization,  
$
\langle \bar B_s|{\cal O}_i|B_s  \rangle= B_i 
\langle \bar B_s|{\cal O}_i|B_s  \rangle^{fac} 
$.
For the operators considered here, we have (e.g.~\cite{beneke:96}) 
(we now use $B$ for the $B_s$ meson and also 
for the bag parameter of the operator $Q$)
\begin{eqnarray}\label{eq:Bdefs}
\langle \bar B|Q|B \rangle&=
& f_B^2 M_B^2 2\left(1+\frac{1}{N_c}\right) B 
 \\
\langle \bar B|Q_S|B  \rangle&=
& -f_B^2 M_B^2 \frac{M_B^2}{(m_b+m_s)^2}
\left(2-\frac{1}{N_c}\right)  B_S
 \\
\langle \bar B|R_2|B  \rangle&=
& -f_B^2 M_B^2 \left(\frac{M_B^2}{m_b^2}-1\right)
\left(1-\frac{1}{N_c}\right)B_2 
 \\
\langle \bar B|R_3|B  \rangle&=
& f_B^2 M_B^2 \left(\frac{M_B^2}{m_b^2}-1\right)
\left(1+\frac{1}{2 N_c}\right) B_3, 
\end{eqnarray}
where $N_c=3$ is the number of colors in QCD
and $f_B$ is the $B_s$ meson semileptonic decay constant.

The dominant theoretical uncertainties in the prediction of 
$\Delta \Gamma=-2\Gamma_{12}$  using the 
heavy-quark expansion are related to the 
hadronic matrix elements of the local operators 
${\cal O}_i \in\{Q,Q_S,R_2,R_3\}$, or equivalently,
the bag parameters $B_i$.
The calculation of the bag parameters 
involves strong interaction dynamics in the infrared 
region and is thus a problem in non-perturbative QCD. 
The ultimate solution can be provided 
by their direct calculation in lattice QCD.  
Results for $Q$ and $Q_S$ are available, 
although not yet completely reliable~\cite{lattice}.  
However, a computation for the operators $R_2$ and 
$R_3$ is completely lacking, and to match the increasing 
precision of the experimental data it is necessary to 
consider deviations from $B_i=1$ even
for these subleading operators~\cite{lenz:06}.

In this paper we use the technique of QCD sum rules to 
provide a first estimate of the bag parameters for the
subleading operators $R_2$ and $R_3$. We focus on the 
calculation of the parameters $\Delta B_i=B_i-1$, which measure
the deviations from the factorization result $B_i=1$.
We limit our analysis to the non-perturbative vacuum condensate
contributions to these quantities. While more sophisticated treatments with
lattice QCD exist for the leading-order operators $Q$ and $Q_S$,
and with QCD sum rules for $Q$, we also include these
operators in our analysis. Studying the full set of operators 
simultaneously helps clarify the general features of  sum rules as
applied to this class of matrix elements. 

Our main finding is that the non-perturbative 
contributions to $\Delta B$ are quite small for each of 
the four operators, no larger than 1-2\%. 
We use a simple analytical analysis based on the HQET limit within 
finite energy sum rules (FESR) to give insight into this result. To
explore corrections to the HQET limit and to provide
error estimates we perform a more thorough numerical 
analysis using Borel sum rules.   The numerical 
results suggest that corrections to the HQET limit 
may be large in some cases.

The paper is organized as follows.
In Section~\ref{sec:sumruleMethod} we describe the technique of sum rules
as applied to our case and introduce some necessary notation.
In Section~\ref{sec:OPE} we describe the calculation of 
operator-product expansion (OPE) expressions for
the Green functions used in the analysis.
Sections~\ref{sec:Sumruleresults} and~\ref{sec:borel} 
contain our sum-rule analysis 
and includes full QCD and the HQET limit in FESR and Borel form.
In Section~\ref{sec:Summing} we give the final results and discuss
the assumptions made and uncertainties involved.
In Section~\ref{sec:Conclusion} we give the summary of the paper.
Some long formulae for the OPE spectral densities are collected in the 
Appendix. 

\section{Sum rule calculation of the bag parameters: \\
the technique}
\label{sec:sumruleMethod}
 
In this section we review the sum-rule method for calculating 
the hadronic matrix elements of the $\Delta B=2$ operators. 
The starting point is the three-point correlator
\begin{equation}
T(p_1,p_2)=i^2 \int d^4 x d^4 y e^{ip_1 x -ip_2 y}
\langle T j(x) {\cal O}(0)j(y)\rangle.
\end{equation}
The operator ${\cal O}\in\{Q,Q_S,R_2,R_3\}$ is a generic four-quark operator 
and the interpolating current $j$ for the $B$-meson can be either
an axial-vector (AV) current or  pseudoscalar (PS) current, defined as
\ba
j^\mu_5 &=&\bar s\gamma^\mu\gamma_5 b \qquad ({\rm AV \, current}) \\
j_5&=&\bar s i\gamma_5 b \hspace {.28cm} \qquad ({\rm PS \, current}) .
\ea
The overlap
of the interpolating currents with $B$-meson states is 
defined through the matrix elements 
\be\label{eq:twopointfns}
\langle 0|\bar s \gamma_\mu \gamma_5 b(0)|\bar B(p)\rangle
=i f_B p_\mu, \qquad 
\langle 0|\bar s i \gamma_5 b(0)|\bar B(p)\rangle
=\frac{f_B M_B^2}{m_b+m_s},
\ee 
where $f_B$ is the semileptonic decay constant of the $B$ meson, $M_B$ is
the $B$-meson mass, $m_b$ is the $b$-quark mass, and $m_s$ is the 
strange-quark mass.
For the axial-vector interpolating current 
the three-point correlator is a 
tensor, and we focus on the scalar function multiplying 
the tensor structure $p_1^\mu p_2^\nu$:
\begin{equation}
T^{\mu\nu}(p_1,p_2)=i^2 \int d^4 x d^4 y e^{ip_1 x -p_2 y}
\langle T j^\mu_5(x) {\cal O}(0)j^\nu_5(y)\rangle
=p_1^\mu p_2^\nu T(p_1,p_2)+\ldots
\end{equation}
where the ellipsis denote other tensor structures such as
$p_1^\mu p_1^\nu$, $p_2^\mu p_2^\nu$,  $p_1^\nu p_2^\mu$ or $g_{\mu\nu}$.   
It is convenient to use the dispersion relation
\begin{equation}
T(p_1,p_2)=\int ds_1 ds_2 \frac{\rho(s_1,s_2,q^2)}{(s_1-p_1^2)(s_2-p_2^2)}
\end{equation}
and work with the spectral density $\rho(s_1,s_2,q^2)$. 
Here $q=p_1-p_2$ and $q^2=0$ at the physical point relevant to the
mixing.  To derive
the sum rules the spectral density is evaluated in two ways:
\begin{enumerate}
\item In a phenomenological hadronic picture. In this case the spectral
density is modeled by a $B$-meson pole plus a continuum contribution.
This yields
\begin{equation}
\rho^{\rm had}_{\rm AV}(s_1,s_2)=\left[f_B^2 \delta(s_1-M_B^2)\delta(s_2-M_B^2)
\langle \bar B| {\cal O}|B \rangle\right]+
\rho^{\rm cont}_{\rm AV}
\end{equation}
for the axial-vector current, and 
\begin{equation}
\rho^{\rm had}_{\rm PS}(s_1,s_2)=\left[\frac{f_B^2 M_B^4}{(m_b+m_s)^2} 
\delta(s_1-M_B^2)\delta(s_2-M_B^2)\langle \bar B| {\cal O}|B
\rangle\right]
+
\rho^{\rm cont}_{\rm PS}
\end{equation}
for the pseudoscalar current.

\item With QCD using the operator-product expansion.  The
resulting spectral densities $\rho^{\rm OPE}_{i}$ are the sum of 
a perturbative contribution and a non-perturbative contribution
involving the vacuum matrix elements of local QCD operators (condensates).  

\end{enumerate}
The  idea of QCD sum rules is to use duality
between the physical spectrum measured in terms of hadrons and the
OPE prediction expressed in terms of quarks and gluons (the
degrees of freedom of the QCD Lagrangian). Duality is implemented by 
comparing integrals of the two spectral densities   
\be\label{eq:dualityintegral}
\int ds_1 ds_2 \, \rho^{{\rm had}}_{i}(s_1,s_2)=
\int ds_1 ds_2 \, \rho^{{\rm OPE}}_{i}(s_1,s_2) .
\ee
It is common practice to model the continuum contribution
to the hadronic spectral density with the theoretical expression
from the OPE.  We choose to match the two expressions
at the point $s_1=s_2=s_0$, so that the integration region $\Delta$ in
the duality integral is the square  $m_b^2<s_i<s_0$ in
the $(s_1,s_2)$ plane. 
One then obtains the sum rules
\begin{eqnarray} 
\label{eq:sumrule1AV}
 f_B^2\langle \bar B| {\cal O}|B \rangle &=& 
\int_\Delta  ds_1 ds_2 \, \rho^{\rm OPE}_{\rm AV}(s_1,s_2)
\qquad ({\rm AV \,current})\\
\frac{M_B^4}{(m_b+m_s)^2}f_B^2
\langle \bar B| {\cal O}|B \rangle &=& 
\label{eq:sumrule1PS}
\int_\Delta  ds_1 ds_2 \,  \rho^{\rm OPE}_{\rm PS}(s_1,s_2) 
\qquad ({\rm PS \,current}) .
\end{eqnarray}
Calculating the OPE expressions 
for the spectral density thus allows for the extraction of the hadronic 
matrix elements $\langle \bar B| {\cal O}_i|B \rangle$, or, equivalently,
the bag parameters $B_i$.  The sum-rule results depend on the
parameter $s_0$ at which the hadronic continuum is 
modeled by the OPE result; we shall discuss different ways of choosing
this parameter later on. 

The sum rules (\ref{eq:sumrule1AV}, \ref{eq:sumrule1PS}) are referred to as 
``finite energy sum rules'' (e.g.~\cite{fesr}).  
It is expected that results obtained with these 
basic sum rules give a reasonable
approximation to a more sophisticated analysis. However,  
it is also useful to consider a different averaging procedure 
in the duality integrals. The most popular technique is the 
Borel sum rule analysis. 
In Borel sum rules one works with duality integrals of moments
of the spectral densities rather than with the spectral densities themselves.
In particular, one compares the derivatives  
$\partial^n/(\partial p^2)^n$ of the
spectral densities for large $n$. In the limit $n\to \infty$ and again
modeling the hadronic continuum with the OPE prediction 
one arrives at the Borel sum rule
\be
\label{eq:sumruleBorel}
 f_B^2\langle \bar B|{\cal O}|B 
\rangle e^{-\frac{M_B^2}{M^2_1}-\frac{M_B^2}{M^2_2}}
= \int_\Delta  ds_1 ds_2\, e^{-\frac{s_1}{M^2_1}-\frac{s_2}{M^2_2}} 
\,\rho^{\rm OPE}_{\rm AV}(s_1,s_2)
\qquad{(\rm AV \,current)}
\ee
and analogously for the pseudoscalar case.
In the Borel sum rule contributions from excited states
are exponentially suppressed. Also, studying the stability of the 
sum rule results under variations of the Borel parameters $M_1$ and
$M_2$ helps assess their reliability. 

The procedure sketched above can be used to compute the bag
parameters directly. However, at the level of the OPE, one can 
identify the contributions to the three-point correlator 
which lead to the value  $B=1$ only~\cite{chet:86,ovch:88}. 
Such contributions can be expressed as the product of two 
color-singlet two-point functions, each
depending on a single momentum. Subtracting this trivial
part from the QCD sum rule allows us to focus on the piece
responsible for deviations from the factorized value.
We thus split  the three-point correlator into two 
pieces according to
\be \label{eq:deltaTdef}
T(p_1,p_2)=T_{\rm fac}(p_1,p_2)+\Delta T(p_1,p_2),
\ee
where the sum rule obtained from the factorized piece $T_{\rm fac}$
yields $B=1$. This factorized part has the explicit form 
\be 
T_{fac}(p_1,p_2)={\rm const}\times \Pi(p_1) \Pi(p_2)
\ee
with the ``const'' and the $\Pi(p_i)$ specific to the operator involved.
For instance, for the operators involving a V-A Dirac structure, one has
\be 
T_{fac}^{\rm AV}(p_1,p_2)=2\left(1+\frac{1}{N_c}\right) 
\Pi^V(p_1) \Pi^V(p_2)
\ee
with 
\be
p^\alpha\Pi^V(p)=i\int dx e^{ipx} 
\langle T j(x) \bar b\gamma^\alpha(1-\gamma_5) s(0)\rangle.
\ee
Using this same notation for the factorizable and non-factorizable
contributions to the spectral densities one finds a sum rule for
$\Delta B=B-1$ directly.  It reads
\be
\label{eq:sumruleBorelDelta}
f_B^2 \Delta B\langle \bar B| {\cal O}|B\rangle^{fac}
e^{-\frac{M_B^2}{M^2_1}-\frac{M_B^2}{M^2_2}}
= \int  ds_1 ds_2  \Delta\rho^{\rm OPE}_{\rm AV}(s_1,s_2)
e^{-\frac{s_1}{M^2_1}-\frac{s_2}{M^2_2}}
\ee
for the Borel sum rule with an ${\rm AV}$ interpolating current 
and analogously for the other cases.

If $\Delta B$ is numerically small compared to 
the factorized value $B=1$
(as one expects from the previous 
analyses~\cite{ovch:88,Reind,narpiv} and the present study 
confirms), then this setup allows for an essential
improvement in precision in comparison with the analysis of 
the $B$ parameter itself. 


\subsection{Sum rules in the HQET limit}
\label{sec:HQETsumrules}
The $\Delta B=2$ operators are identified by evaluating the
transition operator as a series in $1/m_b$, according to the 
heavy-quark expansion. In this treatment, the operators
are defined in terms of QCD fields and contain implicit $m_b$ dependence. 
For processes containing heavy quarks it is advantageous to make this 
$m_b$ dependence explicit by performing
calculations in the formal limit $m_b\to \infty$
using the framework of  HQET. The effective theory 
sets up a systematic expansion in powers of $1/m_b$, and 
separates the perturbative effects occurring at the scale $m_b$ from those
responsible for the hadronic dynamics at the scale 
$\Lambda_{\rm QCD}$.  In addition to our QCD results, we shall
consider our results evaluated in the HQET limit.

To carry out this expansion to a given order in 
$\alpha_s(m_b)$ and $1/m_b$, one must match the 
interpolating currents and the QCD Lagrangian onto 
their HQET expressions, and evaluate the three-point correlator 
in the sum-rule analysis using these effective-theory objects.  
In this paper we shall limit the HQET expansion of a given matrix 
element to leading order in both perturbative
and $1/m_b$ corrections, ignoring even the effects of 
leading-log resummation. To this level of accuracy the matching onto 
HQET is trivial, and can be obtained directly
from the QCD sum-rule expressions  
by making certain substitutions and then expanding in a
series in the large $b$-quark mass. 
On the phenomenological side of the sum rules, this is done 
by writing $M_B=m_b+\bar \Lambda$ and expanding
to leading order in $\bar \Lambda/m_b$.  On the OPE side, this is done 
by writing the spectral variables as $s_i=(m_b+E_i)^2$ and
expanding to leading order in $E_i/m_b$. 

Applying the HQET expansion to the finite energy sum rules 
(\ref{eq:sumrule1AV},\ref{eq:sumrule1PS}) is 
straightforward, and will be discussed in Section \ref{subsec:FESR}. 
In our numerical analysis in Section \ref{sec:borel},
we will also need the HQET limit of 
the QCD Borel sum rule (\ref{eq:sumruleBorel}) (and its PS analogue). 
To obtain the HQET expression,
we choose the Borel parameters $M_1^2=M_2^2=M^2$ 
and define $W=M^2/m_b$.  Performing the HQET expansion yields
\be
\label{eq:sumruleBorelhqet}
 f_B^2\langle \bar B|{\cal O}|B
\rangle_{\rm HQET}
= 4\int_{\tilde \Delta}  dE_1 dE_2\, e^{\frac{(4\bar\Lambda-2E_1-2E_2)}{W}}
\,\tilde\rho^{\rm OPE}_{\rm AV}(E_1, E_2)
\qquad{(\rm AV \,current)}
\ee
where the HQET limit of the matrix elements are defined by 
the expansion of the right-hand side of (\ref{eq:Bdefs}).  In
this case the duality interval $\tilde \Delta$ is given
by $0<E_i<E_0$. The expressions for $\Delta B$ 
are then derived as before.

\section{The OPE  for the three-point correlators}
\label{sec:OPE}
\begin{figure}
\begin{center}
\includegraphics[width=0.65\textwidth]{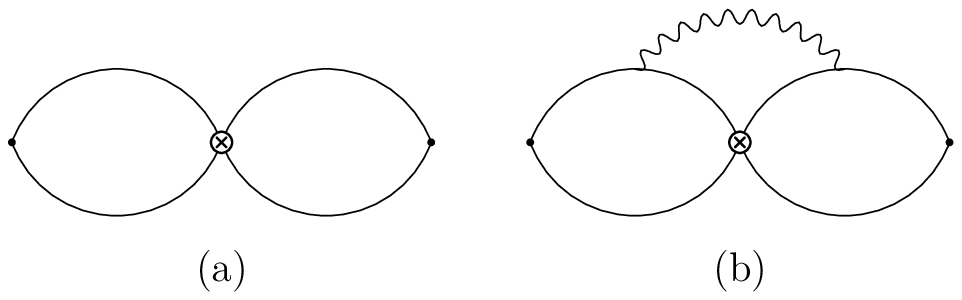}
\end{center}
\vspace{-.5cm}
\caption{The leading-order perturbative contribution to the 
three-point correlator (a), and a non-factorizable
perturbative contribution at next-to-leading-order (b).}
\label{fig:bbmixLO}
\end{figure}

In this section we describe the calculation of the spectral density functions
using the OPE. The leading-order results are given by the bare 
quark loops shown in Figure~\ref{fig:bbmixLO}(a). The cross
denotes the insertion of any one of the four-quark operators 
$Q,Q_S,R_2,R_3$, and the solid dots can be either axial-vector
or pseudoscalar interpolating currents.  The analysis works very much 
the same for each of these eight possible cases.
Corrections to the leading-order
result come from two sources: higher-order perturbative 
corrections and non-perturbative corrections in the form of 
vacuum condensates.  Our focus in this paper is on 
the vacuum condensate contributions, which we 
consider up to dimension six by calculating
the gluon condensate, the mixed quark-gluon
condensate, and the four-quark condensate.

\begin{figure}
\begin{center}
\includegraphics[width=0.95\textwidth]{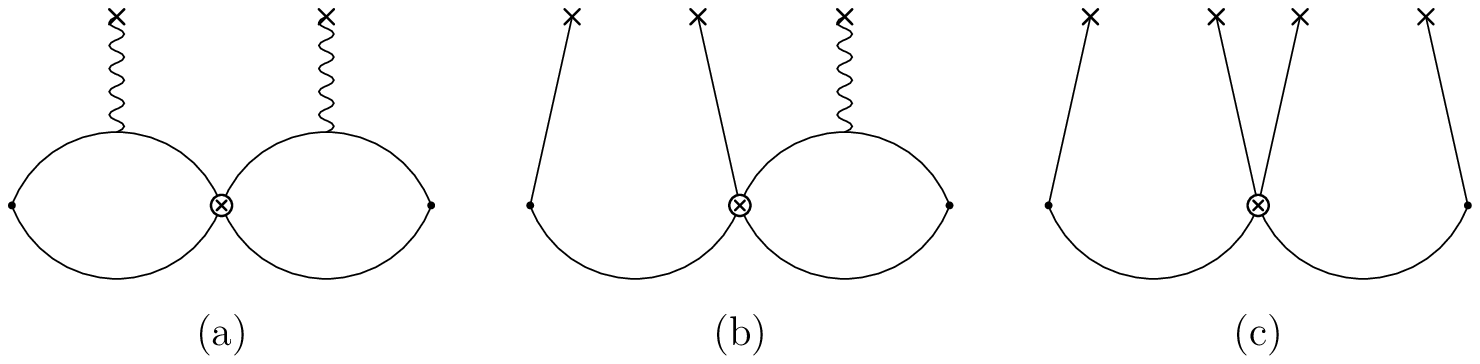}
\end{center}
\vspace{-.35cm}
\caption{\label{fig:condensates}Non-factorizable contributions involving
(a) the $\langle GG \rangle$ condensate, (b) the 
$\langle \bar s G s\rangle$ condensate, and (c) the 
$\langle \bar s  s \bar s s\rangle$ condensate.}
\end{figure}

The leading non-perturbative contributions involve the gluon
condensate, a dimension-four object 
defined through the vacuum matrix element
\be
\label{eq:GGcond}
\langle G_{\mu \nu}^{a}G_{\alpha\beta}^{b} \rangle
=\frac{\delta^{ab}}{12(N_c^2-1)} (g_{\mu\alpha}g_{\nu\beta}-g_{\mu\beta}
g_{\nu\alpha})\langle GG \rangle .
\ee
The non-factorizable corrections proportional to the gluon
condensate are obtained by calculating the diagram 
shown in Figure~\ref{fig:condensates}(a) along
with the three other permutations where the gluons are attached to 
different loops.  Diagrams where the two gluons are attached to 
the same loop are factorizable and hence do not contribute to $\Delta B$.

The calculation is most easily performed using the external-field
method~\cite{Novikov:1983gd}. The advantage of this technique
is that the external 
gluon field can be expressed in terms of the field-strength
tensor according to the relation  
\be
\label{eq:backA}
A^a_\mu(x)=\frac{1}{2}x^\alpha G^a_{\alpha\mu}+O(x^2) .
\ee
This property allows for a direct extraction of 
the gluon condensate contributions
from the diagrams in Fig~\ref{fig:condensates}(a), and also 
simplifies the calculation for the subleading
operators $R_2$ and $R_3$.  Since the operators  $R_2$ and $R_3$ 
are evaluated at 
the point $x=0$, the diagrams where a gluon is emitted from the operator
itself (the cross in the diagrams) vanish, and one need only 
consider derivative couplings, whose evaluation is essentially the 
same as for the leading-order operators $Q$ and $Q_S$.

We next consider the dimension-five contributions. These are 
proportional to the mixed quark-gluon condensate,  which is 
defined through the matrix element 
\be
\label{eq:qqGcond}
\langle\bar s_{\alpha} ig_s G_{\beta\eta}^a t^{a} s_\rho \rangle=
\frac{\left(i\sigma_{\beta\eta}\right)_{\rho\alpha}}{48} 
\langle \bar s G s\rangle.
\ee
The relevant non-factorizable diagrams are shown in
Figure~\ref{fig:condensates}(b).  
As with the gluon condensate, the relation (\ref{eq:backA}) leads to
simplifications for the subleading operators $R_2, R_3$.  Also in 
this case one need not consider gluons emitted from the covariant
derivative; moreover, the external strange-quark fields 
carry vanishing momentum, so derivatives can
only act on the strange-quark field contracted inside the loop.  

Finally, we consider the dimension-six contributions involving
the four-quark condensate.  The relevant non-factorizable diagrams 
are shown in Figure~\ref{fig:condensates}(c).  
These vanish for the subleading operators $R_2$ and
$R_3$, as can be seen by using (\ref{eq:backA}) and then noting
that the derivative terms act on the vacuum fields and thus vanish.
For the leading-order operators $Q$ and $Q_S$ the contributions
involve matrix elements of the form 
$
\langle \bar s \Gamma_1 s \bar s \Gamma_2 s\rangle
$
where the $\Gamma_i$ involve both Dirac and color indices.  
To evaluate these non-factorizable four-quark matrix elements
we use the vacuum saturation approximation, by which the full
matrix element is expressed as 
\begin{equation}\label{eq:4quarkVA}
\langle \bar s \Gamma_1 s \bar s \Gamma_2 s\rangle =\frac{1}{(4N_c)^2}
\left({\rm tr}\Gamma_1\,{\rm tr}\Gamma_2- {\rm tr}\Gamma_1\Gamma_2\right)
\langle {\bar s} s \rangle^2.
\end{equation}
This approximation dates back to the first applications of the 
sum rule method \cite{SVZ}, and since then has been checked 
through numerical analysis in many physical channels.
One particular study for vector-vector and axial-axial 
channels established that the
factorization is accurate within 15-20\%~\cite{chet:pivNC}.
Upon using this approximation for the current correlator, 
we find that non-factorizable contributions from 
the four-quark condensate 
to sum rule for $Q$ and $Q_S$ also vanish.  Details are given
in the appendix.  

We shall limit our OPE analysis to these non-perturbative 
condensates.  To this level of accuracy, the OPE result for the spectral 
density can be written as 
\be
\Delta\rho_i(s_1,s_2)= \Delta\rho_i^{\rm GG}(s_1,s_2)\langle GG \rangle 
+ \Delta \rho_i^{\rm sGs}(s_1,s_2)\langle \bar s G s\rangle + \dots
\ee
for each of the eight cases.  Explicit results for
the $\Delta \rho_i$ can be found in the appendix.
The result for the operator Q with an AV (PS)
interpolating current was first obtained 
in~\cite{ovch:88} (\cite{Reind}), while the others are new.
The ellipsis refers to the corrections not taken into account in our
analysis.  These include contributions from the 
dimension six condensate 
$\langle f^{abc}G^a_{\mu\nu}G^b_{\nu\tau}G^c_{\tau\mu}\rangle$, whose
numerical value is considered to be small~\cite{SVZ}.
An attempt to take into account condensates of operators of
dimension 7 and even 8 was made in ref.~\cite{Reind} for 
the operator $Q$.  We note, however, that the numerical values of 
these condensates are very uncertain and their effects
small, and thus exclude them from the analysis. 

More important are higher-order perturbative corrections. The
next-to-leading order corrections are parametrically on the order of
$\alpha_{s}(m_b)/\pi\sim 0.06$ for 
$\alpha_{s}(m_b)=0.2$.  
Non-factorizable perturbative corrections 
require the evaluation of three-loop diagrams such 
as that shown in Figure~\ref{fig:bbmixLO}(b). These were calculated
in~\cite{bbmixPT} for the leading-order operator $Q$, but are unknown
for the other cases. 

As an example and to introduce notation
we give here the explicit expression for the $Q_S$ operator with 
a pseudoscalar interpolating current: 
\ba
\Delta \rho_{\rm PS}(s_1,s_2) &=& 
\frac{1}{48\pi^2}\langle \frac{\alpha_s}{\pi}GG\rangle
\frac{1}{s_1  s_2}
\left(
\frac{s_1s_2}{2}(6-3z_1-3z_2+z_1 z_2)
+(p_1 p_2)^2 z_1z_2 \right) \nonumber \\
&&+\frac{1}{16\pi^2}\langle \bar sG s\rangle
 m_b \left((-2+z_1)\delta(s_2-m_b^2)
+(-2+z_2)\delta(s_1-m_b^2)
\right)
\ea
Here $z_i=m_b^2/s_i$, and  $\delta(s_i-m_b^2)$ is the Dirac $\delta$ function. 
At the physical point $q^2=(p_1- p_2)^2=0$ the scalar product 
$(p_1 p_2)$ should be understood as $(p_1 p_2)=(s_1+s_2)/2$.

We also need the HQET expansion of the spectral density, which
we obtain by using  $s_i=(m_b+E_i)^2$ and expanding to lowest order 
in $E_i/m_b$.  In this case this limit reads 
\be 
\Delta \rho_{\rm PS}^{\rm HQET}(E_1,E_2)=
\frac{1}{48\pi^2}\frac{1}{4\pi^2}
\bigg[\frac{3}{2}\langle g_s^2 GG\rangle 
-6 \pi^2 \langle \bar s G s\rangle
\left(\delta(E_1)+\delta(E_2)\right)\bigg] .
\ee

\section{The bag parameters from finite energy sum rules}
\label{sec:Sumruleresults}

In this section we present the sum-rule results for the $\Delta B_i$ using
the finite energy sum rules (\ref{eq:sumrule1AV}, \ref{eq:sumrule1PS}) 
evaluated at leading order in the HQET approximation.  We first give simple 
analytical expressions for the $\Delta B_i$, obtained by relating the sum-rule
parameter $s_0$ to the $B$-meson decay constant $f_B$, thereby eliminating
one parameter. Upon inserting numerical values it becomes clear that 
$\Delta B$ is suppressed by a small scale ratio, 
independent of the particular operator being considered. 

\subsection{The choice of duality interval}
The sum-rule results for the $\Delta B_i$ depend on the choice of 
the parameter $s_0$ defining  the upper limit in the duality integrals in
(\ref{eq:sumrule1AV}), (\ref{eq:sumruleBorelDelta}).  For the hadronic part 
the best accuracy is obtained by considering small values of $s_0$
for which saturation by the ground state is a justified approximation.
The OPE side, on the other hand, is best suited for inclusive 
quantities for which perturbation
theory is valid. The quantity $s_0$ must be chosen in such a way
as to balance between these two cases, and the exact value to use
is thus a matter of judgement. A useful guide for determining
its value is to use QCD sum rules for the matrix elements 
(\ref{eq:twopointfns}) to express $s_0$ in terms of the decay constant $f_B$.  
This makes the finite energy sum rule analysis of the three-point correlator 
parametrically free and the analytical results simple,
allowing us to discuss qualitative features which are less transparent in a
purely numerical analysis.

The two-point sum rule for the decay constant $f_B$ is  
obtained in the standard way. One evaluates the spectral density
for the two-point function in both a phenomenological hadronic picture and in 
the OPE.  Equating the integrals of the two spectral densities over a duality
interval gives a result for the decay constant $f_B$. We calculate the OPE
spectral density 
by evaluating the two-point function in its crudest approximation, 
including only the bare quark loop. For the two-point function of axial vector 
currents we have
\be
\rho^{{\rm OPE}}_{\rm AV}(s)=\frac{1}{4\pi^2} (1-z)^2(1+2z), 
\qquad z=m_b^2/s.
\ee
For the phenomenological spectral density we have
\be
\rho^{{\rm PH}}_{\rm AV}(s)=f_B^2\delta(s-M_B^2).
\ee
Equating the two expressions as in (\ref{eq:dualityintegral}) 
(local duality finite energy sum rules~\cite{locduality}) yields
\be
\label{s0AVduality} (2\pi f_B)^2=s_0(1-z_0)^3 \qquad ({\rm AV\, current})
\ee
$z_0=m_b^2/s_0$. We see that the duality 
interval parameter $s_0$ can be expressed
through $f_B$ and $m_b$.  We rewrite the expression~(\ref{s0AVduality}) in 
a form suitable for HQET by substituting
$s_0=(m_b+E_0)^2$ and expanding 
in the ratio $E_0/m_b\ll 1$. Retaining the leading term of the expansion
we find an equation relating the HQET sum-rule parameter $E_0$ with the 
physical quantity $f_B$:
\be\label{eq:E0AV}
(2\pi f_B)^2 \approx \frac{8}{m_b}E_0^3 
\qquad{\rm (AV\,current)}.
\ee
For $f_B=240~\MeV$ and $m_b=4.8~\GeV$ 
one finds $E_0=1.1~\GeV$.
Repeating the analysis for the pseudoscalar interpolating current,
where (neglecting the strange-quark mass)
\be
\rho^{\rm OPE}_{\rm PS}
=\frac{3}{8\pi^2} s\left(1-\frac{m_b^2}{s}\right)^2
\ee
we have in the HQET limit 
\be\label{eq:E0PS}
(2\pi f_B)^2\approx \frac{4}{m_b}E_0^3
\qquad{(\rm PS\, current)}
\ee
which gives $E_0=1.4~\GeV$. Thus, the numerical value of the 
duality interval fluctuates depending on the channel chosen for
its determination. 
At any rate the results are consistent with the
general expectation that the scale of duality in hadronic physics 
is about $1~\GeV$. 

This idea of determining the value of the duality interval 
from two-point sum rules works well quantitatively also for 
light quarks. Indeed, by comparison, for light 
$u$-, $d$-quarks one finds the relation 
$(2\pi f_\pi)^2=s_0$,  which gives 
$s_0=0.7~\GeV^2$ for $f_\pi=130~\MeV$. This is the actual duality
parameter for sum rules in the axial-vector channel of light 
mesons~\cite{radushkin}.

The relations (\ref{eq:E0AV}) and (\ref{eq:E0PS})
allow for a simple parameter-free analysis in 
the HQET limit. They show the correct scaling for the semileptonic 
decay constant with the heavy quark mass, $f_B\sim 1/\sqrt{m_b}$, 
and upon using them in the sum rules 
the explicit results for the bag parameters become 
independent of $m_b$, as appropriate for hadronic quantities.
For a quantitative comparison with full QCD 
higher-order corrections in $E_0/m_b$ 
are important numerically, as the expansion parameter 
$E_0/m_b \approx 0.2$ is not very small.  We see this further in
our analysis with Borel sum rules.

\subsection{Finite energy sum rules in HQET}
\label{subsec:FESR}
In this section we present the analysis using the finite energy 
sum rules (\ref{eq:sumrule1AV}, \ref{eq:sumrule1PS}) expanded to leading
order in HQET. We work at leading order in $\alpha_s$ and ignore
even leading-log resummation. At this level of precision
the HQET approximation can be obtained by first
evaluating matrix elements in full QCD and then expanding as 
described in Section \ref{sec:HQETsumrules}.  To evaluate the 
phenomenological side of the sum rules we use the explicit expressions
(\ref{eq:Bdefs}), and to evaluate the OPE side
we use the HQET results from the appendix. 

\subsubsection{Leading order operators $Q$ and $Q_S$}
We start our analysis with the leading-order operators $Q$ and $Q_S$,
for which we describe the procedure in some detail.

\paragraph{Operator $Q$ with axial vector interpolating current:}

On the phenomenological side of the sum rule (\ref{eq:sumrule1AV})
we have after subtracting the factorized contribution (cf.
eq.~\ref{eq:sumruleBorelDelta})
\be
\label{QAVHQETphside}
I^{PH}=\frac{8}{3}\Delta B f_B^4 M_B^2 \approx \frac{8}{3}\Delta B f_B^4 m_b^2,
\ee 
where in the second equality we used the HQET limit.  To evaluate
the OPE side in the same limit we use $s_i = (m_b+ E_i)^2$ in the
QCD spectral density from the Appendix and expand
to leading order in $E_i/m_b$, leaving 
\ba
\Delta \rho^{OPE}(s_1,s_2)&=&
\frac{1}{48\pi^2}\langle \frac{\alpha_s}{\pi}GG\rangle
\frac{(p_1,p_2)}{s_1  s_2} 2 z_1 z_2 (-3+z_1+z_2-2 z_1 z_2)\nn \\
&\approx&
\frac{1}{48\pi^2}\frac{\langle g_s^2G^2\rangle}{4\pi^2}
\frac{1}{m_b^2}(-6) .
\ea
Performing the integration on the OPE side
we arrive at the sum rule 
\be
\label{eq:sumrulHQETVA}
\frac{8}{3}\Delta B (2\pi f_B)^4=-2\langle g_s^2 G^2\rangle
\frac{E_0^2}{m_b^2} .
\ee
Using  (\ref{eq:E0AV}) to trade $(2\pi f_B)^4$
for $E_0$ we find the simple result
\be
\Delta B=-\frac{3}{256}\frac{\langle g_s^2 G^2\rangle}{E_0^4} .
\ee
The result for the non-perturbative 
bag parameter is independent of $m_b$, as it should be in the HQET
limit, where dynamical quantities depend on soft physics only.
This fact can be noticed already from (\ref{eq:sumrulHQETVA}) 
by using the scaling relation $f_B\sim 1/\sqrt{m_b}$ deduced
from (\ref{eq:E0AV}).
Taking the value of the gluon condensate as  
$\langle g_s^2 G^2\rangle=0.48~\GeV^4=(0.83~\GeV)^4$~\cite{SVZ}
we have 
\be \label{eq:dBQAV}
\Delta B=-0.006, 
\ee
at $E_0=1~\GeV$, which shows 
that the non-factorizable contribution to the matrix 
element is tiny. 

Examining the expressions for $\Delta B$, one sees that 
it is the suppression by the combination of variables 
$(2\pi f_B)^4 m_b^2/E_0^2=64 E_0^4=(3.3~\GeV)^4$
which leads to this result.  This combination 
does not scale with $m_b$ in the HQET limit and
$f_B$ is further enhanced by $N_c^{1/2}$ in the large-$N_c$ limit.
Since the scale of the gluon condensate is given
by $(0.83~\GeV)^4$, the result for  $\Delta B$ is 
proportional to the fourth power of a small number.
In the absence of any accidental numerical enhancement
of the coefficients, which we do not see, the ``natural''
size of the deviations from factorization is  extremely small. 

The answer (\ref{eq:dBQAV})
is the leading-order HQET result.  To get a feel
for the size of the subleading terms, we list the next few terms
in the expansion of the OPE spectral density:
\begin{eqnarray}
\Delta B &=&-\frac{3}{256}\frac{\langle g_s^2 G^2\rangle}{E_0^4}
\left[1-\frac{11}{3}\frac{E_0}{m_b}+10 \left(\frac{E_0}{m_b}\right)^2
-\frac{215}{9} \left(\frac{E_0}{m_b}\right)^3+\dots\right] \nonumber \\
&=&-\frac{3}{256}\frac{\langle g_s^2 G^2\rangle}{E_0^4}
\left[1-0.8 + 0.4  -0.2 + \dots \right]\nonumber \\
&=& -\frac{3}{256}\frac{\langle g_s^2 G^2\rangle}{E_0^4}
\left(0.5 \right),
\end{eqnarray}
where we used $m_b=4.8~\GeV$ and to obtain the last line
we evaluated the full QCD result.  This shows that the subleading
terms are not small, and that keeping only 
the leading-order term misses  the full QCD result by a 
factor of two, at least at $E_0=1~\GeV$.  Given the small
size of $\Delta B$ the factor of two is numerically irrelevant, 
and is actually within the uncertainties of the analysis.  
We return to this point in Section \ref{sec:Summing}, 
using the subleading operator $R_2$ as an additional example.

\paragraph{Operator $Q$ with pseudoscalar interpolating current:}
We can repeat the computation using a pseudoscalar interpolating
current. At leading order in $1/m_b$ we neglect $m_s$ and expand
as before, finding
\be
\frac{8}{3}\Delta B (2\pi f_B)^4= 
\frac{1}{m_b^2}\left(
-\langle g_s^2 G^2\rangle E_0^2
+8\pi^2\langle \bar s G s \rangle E_0
\right).
\ee
The mixed quark-gluon condensate is parameterized as
$\langle \bar s G s \rangle= m_0^2 \langle \bar s  s \rangle$. 
For numerical evaluation we use  
$m_0^2=0.8~\GeV^2$~\cite{ioffe:m02,ovchpiv:m02} and  
$\langle \bar s  s \rangle = 0.8 \langle \bar u  u 
\rangle$~\cite{narison:gamma,ioffe:gamma}.
For the light quark condensate $\langle \bar u  u \rangle$
we take $\langle \bar u  u \rangle = (-0.24~\GeV)^3$.
A convenient normalization for the mixed quark-gluon condensate 
is
\be
\pi^2\langle \bar s G s \rangle
=-0.1~\GeV^5=(-0.63~\GeV)^5,
\ee
which is of the same order of magnitude as $\langle g_s^2 G^2\rangle$
and is really given by the hadronic scale of $1~\GeV$ (or 
say by the $\rho$-meson mass 
$m_\rho=770~\MeV$). We see that all dimensionful quantities are on 
the order of the fundamental QCD scale of $1~\GeV$ as expected.

Using (\ref{eq:E0AV}) to eliminate $f_B$ one finds
\be
\Delta B=-\frac{3}{256}
\left(\frac{\langle g_s^2 G^2\rangle}{2E_0^4}
-\frac{4\pi^2\langle \bar s G s \rangle}{E_0^5} \right)=
-\frac{3}{256}\left(0.24+0.4\right)=-0.008
\ee
at $E_0=1~\GeV$. 
The result is more or less the same as with the axial-vector 
interpolating current but the structure of contributions changed.
The gluon condensate contributes less and mixed quark-gluon gives 
a contribution (it was zero for the axial-vector case).

Using $E_0^{PS}=1.4~\GeV$ from (\ref{eq:E0PS}) for the pseudoscalar channel
one finds 

\be
\Delta B=-\frac{3}{64}
\left(\frac{\langle g_s^2 G^2\rangle}{2(E_0^{PS})^4}
-\frac{4\pi^2\langle \bar s G s \rangle}{(E_0^{PS})^5} \right)=
-\frac{3}{64}\left(0.062+0.074\right)=-0.006
\ee
which coincides with the previous result from the AV current.  
Even though it is more proper to use the result 
(\ref{eq:E0PS}) for the PS channel, 
for the remaining operators  we shall use
(\ref{eq:E0AV}) in both the AV and PS channel; the differences
are very small.

\paragraph{Operator $Q_S$ with axial vector interpolating current:}
Repeating the analysis for $Q_S$ with an axial-vector current
we find the sum rule
\be
-\frac{5}{3}\Delta B_S (2\pi f_B)^4= 
\frac{1}{m_b^2}\left(
\frac{1}{6}\langle g_s^2 G^2\rangle E_0^2
-4\pi^2\langle \bar s G s \rangle E_0
\right)
\ee
and 
\be
\Delta B_S=-\frac{1}{640}
\left(\frac{\langle g_s^2 G^2\rangle}{E_0^4}
-24\pi^2\frac{\langle \bar s G s \rangle}{E_0^5} \right)=
-\frac{1}{640}\left(0.48+2.4\right)=-0.005
\ee
again a very small number.
Note that the result is dominated by the contribution of the mixed
quark-gluon condensate, even though it is formally subleading compared
to the gluon condensate.  This is a general situation, as 
contributions from the gluon condensate are often small numerically.

\paragraph{Operator $Q_S$ with pseudoscalar interpolating current:}
Repeating for the pseudoscalar current we have
\be
\Delta B_S =-\frac{3}{640}
\left(\frac{\langle g_s^2 G^2\rangle}{E_0^4}
-8\pi^2 \frac{\langle \bar s G s \rangle}{E_0^5} \right)=
-\frac{3}{640}\left(0.48+0.8\right)=-0.006 .
\ee
The coefficients of the gluon and mixed quark-gluon condensates changed,
but their sum is very close to that obtained with the axial-vector current.

We conclude that the deviation from factorization is tiny 
just because of the scales involved. No surprisingly 
big numbers or drastic cancellations occurred in
the analysis.

\subsubsection{Subleading operators $R_2$ and $R_3$}
The analysis is essentially unchanged for the subleading operators
$R_2$ and $R_3$. The new feature is the appearance of the
parameter $\bar \Lambda=M_B-m_b$ even at leading-order in 
the HQET expansion (numerically $M_B=5367.5\pm 1.8~\MeV$~\cite{PDG}). 
It enters through the expansion of the matrix 
elements (\ref{eq:Bdefs}), which read
\begin{eqnarray}
\langle \bar B|R_2|B  \rangle_{\rm HQET}&=
& -f_B^2 m_b^2 \left(2\bar\Lambda m_b \right)
\left(1-\frac{1}{N_c}\right)B_2  \\
\langle \bar B|R_3|B  \rangle_{\rm HQET}&=
& f_B^2 m_b^2 \left(2 \bar\Lambda m_b\right)
\left(1+\frac{1}{2 N_c}\right) B_3.
\end{eqnarray}
This $\bar\Lambda/m_b$ power suppression of the matrix 
elements on the phenomenological 
side of the sum rules is compensated by an $E_0/m_b$ suppression from the 
OPE spectral densities.  Then, up to a factor of $E_0/2\bar \Lambda\sim1$,
the magnitude of $\Delta B$ for the subleading operators is 
fixed by the same scale ratios as before, and as with the leading-order case 
there are no large deviations from factorization.

\paragraph{Operator $R_2$ with axial vector interpolating current:}

\be
-2 \bar\Lambda m_b (2\pi f_B)^4 \frac{2}{3}\Delta B_2=\frac{1}{m_b^2}
\left(
\langle g_s^2 G^2\rangle m_b E_0^3 \left(-\frac{3}{2}\right)
-2\pi^2 \langle \bar s G s\rangle m_b E_0^2 \right)
\ee
Notice that for this subleading operator the phenomenological
and OPE sides of the sum rule are suppressed by the hadronic
scales $\bar \Lambda$ and $E_0$ respectively.  
This is explicit in the HQET expressions
but not in the QCD ones. Using (\ref{eq:E0AV}) and 
taking $\bar \Lambda=0.5~\GeV$ 
\be
\Delta B_2=\frac{E_0}{2\bar \Lambda}\frac{9}{256}
\left(\frac{\langle g_s^2 G^2\rangle}{E_0^4}
+\frac{4\pi^2\langle \bar s G s \rangle}{3 E_0^5} \right)=
\frac{9}{256}\left(0.48-0.13\right)=0.012 .
\ee
We see that $\Delta B_2$ is again very small, although this time
it is a {\it positive} number instead of a negative one.

\paragraph{Operator $R_2$ with pseudoscalar interpolating current:}
\be
\Delta B_2=\frac{E_0}{2\bar \Lambda}\frac{3}{256}
\left(\frac{\langle g_s^2 G^2\rangle}{E_0^4}
-\frac{4\pi^2\langle \bar s G s \rangle}{E_0^5} \right)=
\frac{3}{256}\left(0.48+0.4\right)=0.010 .
\ee

\paragraph{Operator $R_3$ with the axial vector interpolating current:}
\be
\frac{7}{6}\Delta B_3=\frac{E_0}{2\bar \Lambda}\frac{1}{64}
\left(\frac{\langle g_s^2 G^2\rangle}{E_0^4}\left(\frac{1}{6}\right)
-\frac{2\pi^2\langle \bar s G s \rangle}{E_0^5} \right)=
\frac{1}{64}\left(0.08+0.2\right)=0.004 .
\ee
and 
\be
\Delta B_3=0.004
\ee

\paragraph{Operator $R_3$ with pseudoscalar interpolating current:}
\be
 \frac{7}{6}\Delta B_3=\frac{E_0}{2\bar \Lambda}\frac{1}{64}
\left(\frac{\langle g_s^2 G^2\rangle}{E_0^4}\left(\frac{1}{4}\right)
-\frac{\pi^2\langle \bar s G s \rangle}{E_0^5} \right)=
\frac{1}{64}\left(0.12+0.1\right)=0.003 
\ee
and 
\be
\Delta B_3=0.003 .
\ee

We can summarize by saying that the finite energy sum rules
within the HQET approximation suggest that factorization
is perfectly precise, if only non-perturbative condensate
effects are taken into account. 
The bag parameter $B_2$ has the largest violation of 
factorization, but it is still very small in absolute
terms, approximately 1\%.

\section{The bag parameters from Borel sum rules}
\label{sec:borel}
We have seen in the previous section that the deviations from 
factorization for both the leading and subleading operators 
are very small.  
In this section we perform a more thorough numerical
analysis using Borel sum rules.  This serves to
confirm these results and to show that it is possible to impose very 
conservative error estimates without
altering this conclusion. We also use the numerics to compare 
the HQET and full QCD results.

\subsection{Borel sum rules in full QCD}
\label{sec:borelfull}

\begin{figure}[t]
\begin{center}
\includegraphics[width=.9\textwidth]{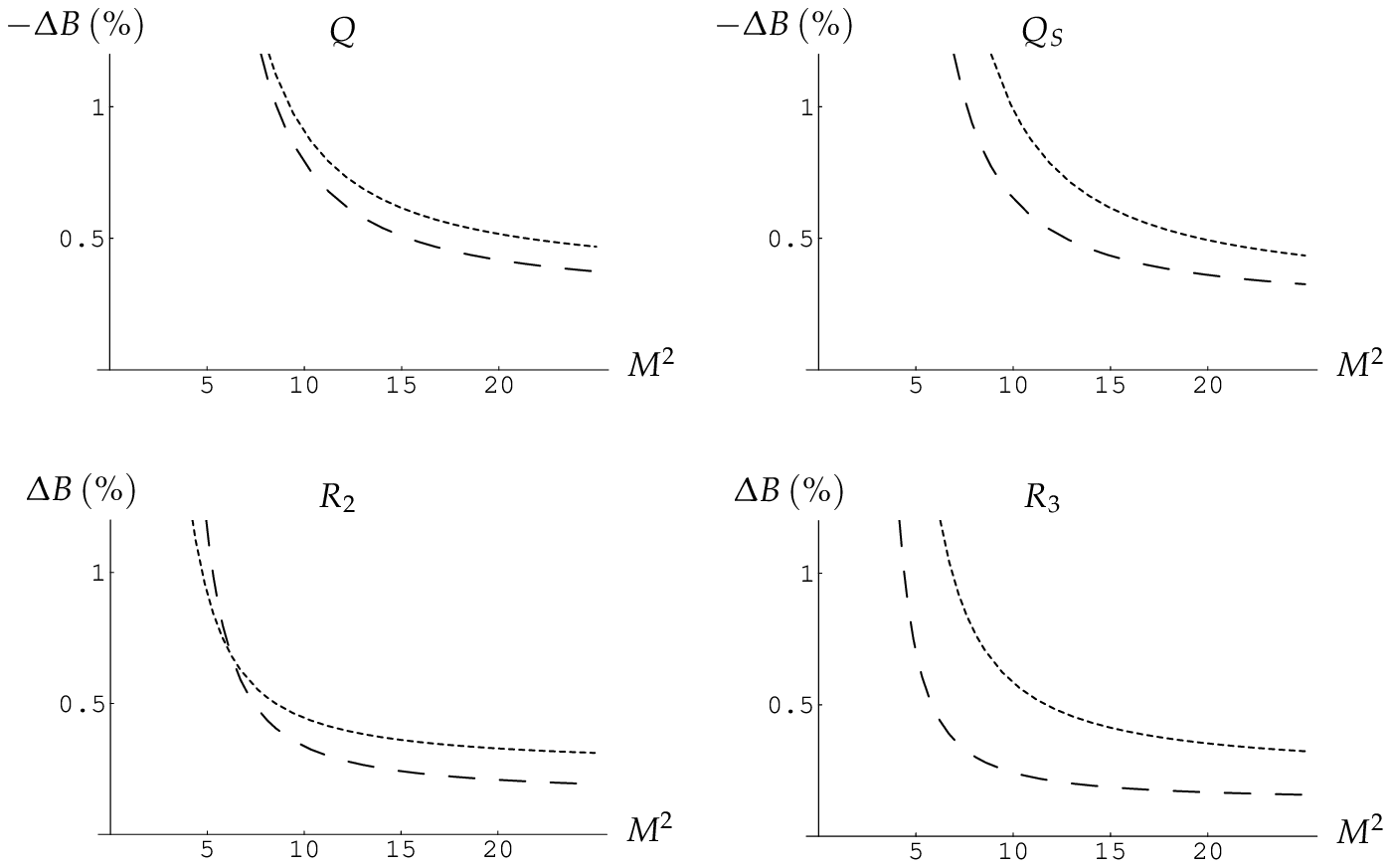}
\end{center}
\vspace{-0.5cm}
\caption{\label{fig:BorelQCD}
Plot of $-\Delta B$ vs. $M^2\,({\rm GeV}^2)$ for 
the leading-order operators $Q$ and $Q_S$
and  $\Delta B$ for the subleading operators $R_2$ and $R_3$
with the Borel sum rules in full QCD. 
The short-dashed lines are obtained using an axial vector
interpolating current, and the
long-dashed lines using a pseudoscalar current.
The parameter values are given by
$m_b=4.2$~GeV,  $f_{B_s}=240$~MeV, $s_0=36~\GeV^2$, and $m_s$
as explained in the text.}
\end{figure}

The Borel sum rules in full QCD are evaluated according
to (\ref{eq:sumruleBorel}) and the analogous expression
for the pseudoscalar interpolating current.
Although it is possible to evaluate the Borel integrals 
analytically, the results are quite lengthy and we do not 
need them for this purely numerical analysis.

To evaluate the sum rules, we must first give numerical values
for the QCD parameters $f_{B_s}$, $m_b$, and $m_s$.  For the decay constant
we choose $f_{B_s}=240~\MeV$ as the default value.   
For the $b$-quark mass one can take the pole mass or the $\MSbar$ mass.  
The pole mass is $m_b^{pole}=4.8~\GeV$ while the $\MSbar$ value is 
$m_b^{\MSbar}=4.2~\GeV$~\cite{Penin:1999kx,Pineda:2006gx}. For 
the full QCD analysis the $\MSbar$ mass is more 
appropriate. However, since we are working to lowest 
order in $\alpha_s$, we cannot distinguish these two
quark-mass definitions, and the difference can be accounted
for as an additional uncertainty in $\Delta B$.
This difference would be under control if $\alpha_s$ corrections were
taken into account.

The strange-quark mass appears on the OPE side of the sum rules for all
channels, and in the phenomenological side for the case of the pseudoscalar
interpolating current. We have seen that $\Delta B$ is extremely small in all
cases, and the effects of a non-zero strange quark mass do little to alter
this. We choose to keep it non-zero on the phenomenological side for the
leading-order operators $Q$ and $Q_S$, using
$m_s=100~\MeV$.  Keeping it non-zero
in the OPE spectral densities 
complicates the analytical expressions without changing the final 
results in a significant way.  For the subleading operators it is 
consistent to set it to zero at this order in the heavy-quark expansion.

The results for $\Delta B$ vs.~the Borel parameter 
$M^2$ for each operator are shown in  Figure~\ref{fig:BorelQCD}. The
two lines in each plot are obtained by using an axial-vector
and pseudoscalar interpolating current. All results have a reasonable 
stability region in $M^2$ at $10~\GeV^2<M^2<20~\GeV^2$.
There is some dependence on the choice of interpolating current, which as we 
will see later is within the uncertainties of the analysis.
We note again that $\Delta B$ is {\it positive} for the 
subleading operators $R_2$ and $R_3$, whereas it is
{\it negative} for the leading operators $Q$ and $Q_S$.

\subsection{Borel sum rules in HQET}
\label{sec:borelHQET}

\begin{figure}[t]
\begin{center}
\includegraphics[width=.9\textwidth]{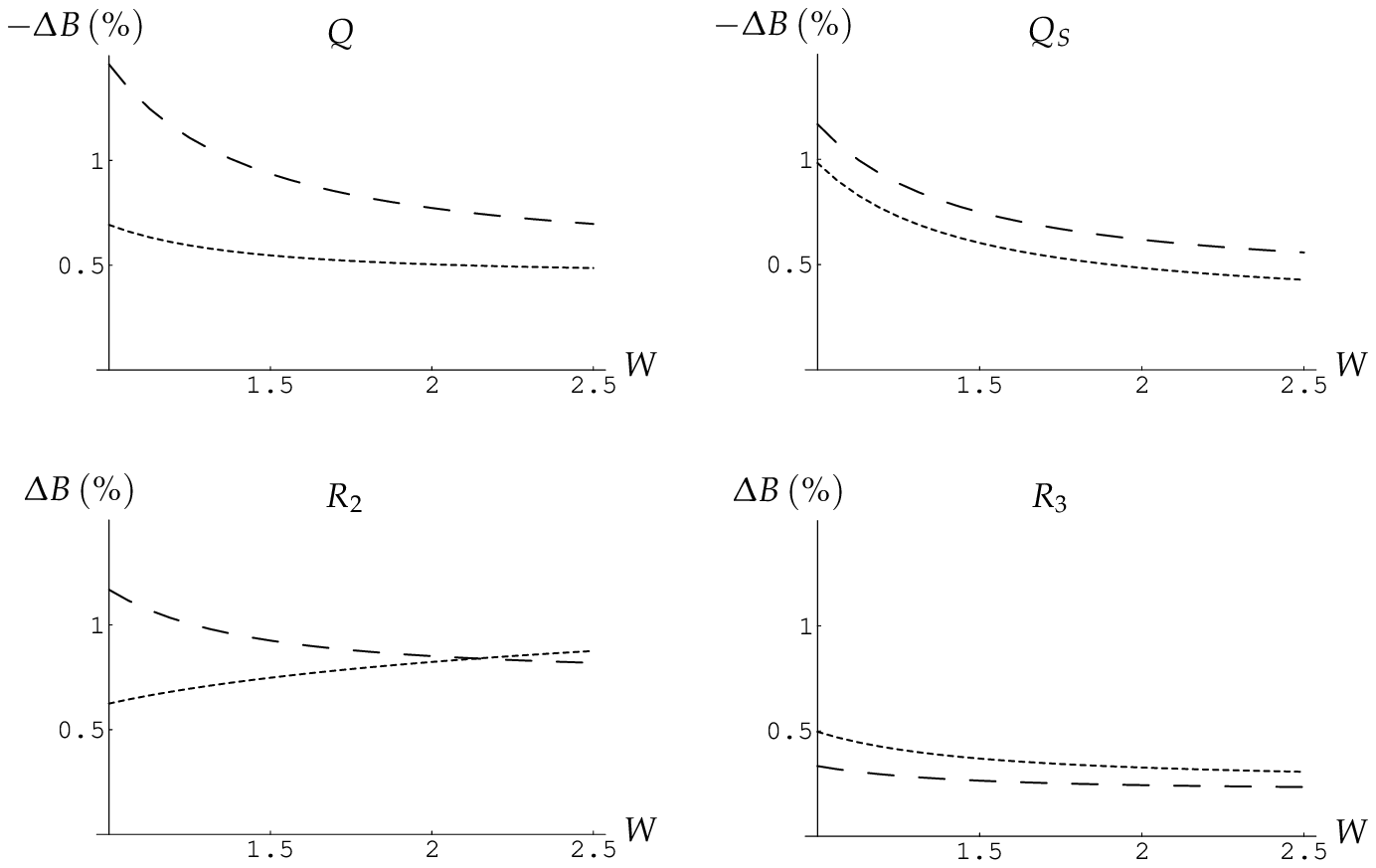}
\end{center}
\vspace{-0.5cm}
\caption{\label{fig:BorelHQET}
Plots of $(-)\Delta B$ vs. $W~(\GeV)$
obtained with the Borel sum rules in HQET. The short-dashed lines are
obtained using an axial vector
interpolating current, and the long-dashed lines using
a pseudoscalar current.
We take $\bar \Lambda=600$~MeV, $m_b=4.8~\GeV$, $f_{B_s}=240$~MeV, and
$2E_0=2.5~\GeV$. }
\end{figure}

The Borel sum rules in HQET are performed according to
(\ref{eq:sumruleBorelhqet}) and analogously
for the PS interpolating current.  We focus on numerical 
results, although the analytical results for the Borel 
integrals are very simple. In fact, they reduce
to those from the finite energy sum rules in the limit $W\to \infty$.
In contrast to our treatment of the finite energy sum rules, however,
in our numerical studies we treat $E_0$ and
$f_B^2 m_b$ as  free parameters.  We again use 
$f_{B_s}=240~\MeV$ as the default
value. While in the QCD calculation the $\MSbar$ mass was more natural,
in HQET the pole mass appears in the construction
of the effective theory and is more natural.
We use $m_b^{pole}=4.8~\GeV$.

The results for $\Delta B$ vs.~the HQET Borel parameter 
$W$ for each operator are shown in  Figure~\ref{fig:BorelHQET}.  
The plots are stable in the region $1~\GeV<W<2.5~\GeV$, 
which is rather typical for Borel sum rules in HQET. 
The values of $\Delta B$ in the stability range are close 
to those in the QCD plots in  Figure~\ref{fig:BorelQCD}.
The one noticeable exception is $R_2$, where the HQET values are 
about twice as large as the QCD ones.  We comment further 
on this in the next section.
\begin{table}
\begin{center}
\begin{tabular}{||c|c|c||}
\hline\hline
Operator & $\Delta B (\%)$ QCD &$\Delta B (\%)$ HQET  \\
\hline
$Q$
& $-0.6 \pm 0.5$
& $-0.6 \pm 0.5$  \\
$Q_S$
& $-0.5 \pm 0.4$
& $-0.6 \pm 0.4$ \\
\hline
$R_2$
& $0.3 \pm 0.3$
& $0.8 \pm 0.7$  \\
$R_3$
&  $0.3 \pm 0.2$
&  $0.3 \pm 0.2$ \\
\hline\hline
\end{tabular}
\end{center}
\caption{\label{tab:results} A summary of the results.}
\end{table}


\begin{figure}[t]
\begin{center}
\includegraphics[width=.8\textwidth]{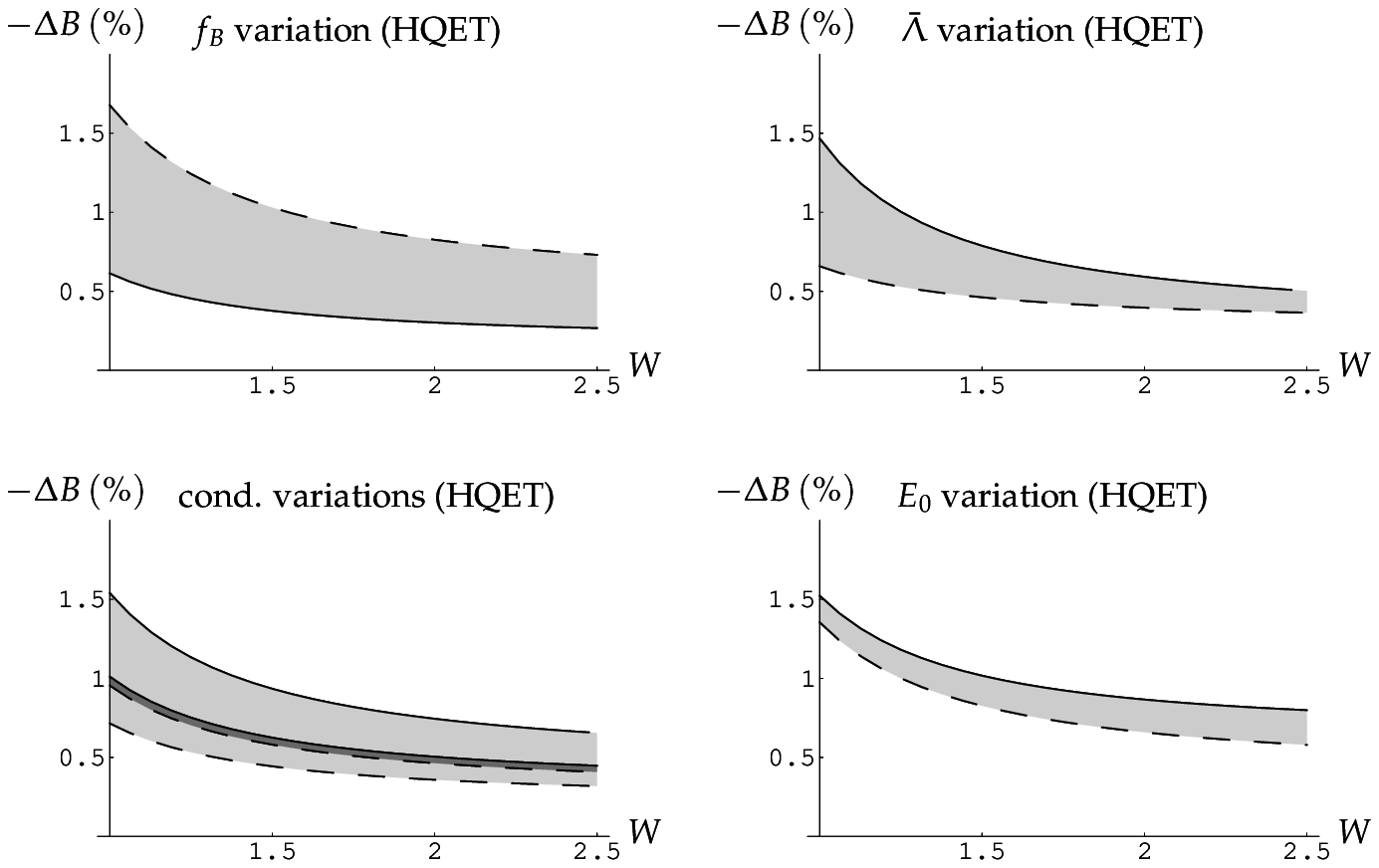}
\end{center}
\vspace{-0.5cm}
\caption{\label{fig:ParVar}
Plots showing $-\Delta B$ vs. $W~(\GeV)$ for the case
of the $Q_S$ operator, axial-vector current,
using a range of parameter values in the HQET
sum rule. The plots are made by varying the parameters 
as explained in the text.
For the condensate variations,  the
dark-gray band corresponds to the gluon condensate
and the larger light-gray band to the quark-gluon condensate variation.
In each case, the dashed line corresponds to the
lower border of the parameter range (e.g. $f_{B_s}=210$~MeV) and the solid
line to the upper border (e.g. $f_{B_s}=270$~MeV).}
\end{figure}

\section{Final  results and discussion}
\label{sec:Summing}
We now present our final numerical results and 
estimate the associated uncertainties.
The results are summarized in Table \ref{tab:results}.

To obtain the table entries for full QCD, 
we fix the Borel parameter at $M^2=15~\GeV^2$ 
and vary the other parameters in the ranges
$$ {\underline{\rm QCD}}$$
$$210~\MeV < f_{B_s} <270~\MeV $$
$$ 4~\GeV < m_b <4.4~\GeV $$
$$ 32~\GeV^2 < s_0 < 40~\GeV^2 $$
where the default values lie in the
center of the above ranges. We also vary the condensates about 
their default values by $\pm 30\%$. For a
given case, we find upper and lower values of $\Delta B$ to 
identify the error ranges.  
For the $f_{B_s}, m_b$ and $s_0$ variations the ranges are 
asymmetric; in those cases we use the larger deviation in
the error analysis.  Finally, we add the uncertainties
from each of the five variations in quadrature, and
average the results from the axial-vector and 
pseudoscalar interpolating currents to  obtain the 
results quoted in the table.

The procedure is the same for the HQET sum rules, although the
set of parameters is different.  This time
we fix the Borel parameter at $W=2~\GeV$ and 
the $b$-quark mass at $m_b=4.8~\GeV$,
and vary the other parameters in the ranges
$$ {\underline{\rm HQET}}$$
$$ 210~\MeV < f_{B_s} <270~\MeV $$
$$ 500~\MeV < \bar \Lambda <700~\MeV$$
$$ 1~\GeV <  E_0 < 1.5~\GeV$$ 
where the default values lie in the center of the above ranges. The 
condensates are again varied by $\pm 30\%$ about their default 
values. The final table entries are  obtained as for the QCD case.

To illustrate the uncertainty associated with each parameter variation, we 
choose as an example the $Q_S$ operator with an axial-vector 
interpolating current in HQET. The range of $\Delta B$ associated with
each variation is represented by the gray bands in Figure \ref{fig:ParVar}.
It is seen that the largest errors are
associated with the value of the decay constant $f_B$.  This is not
surprising, since the explicit results scale as $1/f_B^4$.  At the default
value $W=2$ the dependence on $\bar\Lambda$ and $E_0$ is moderate.  
The results depend linearly on the condensates and at $W=2$ 
the uncertainty due to the condensates is comparable with that 
due to $f_B$.

In all cases except for $R_2$, our central values for 
$\Delta B$  in QCD and HQET turned out to be (nearly) equal.
However, in interpreting this result, one should be 
clear that not only the bag parameters, but 
also the QCD parameters $f_B$ and $M_B$ have an expansion in $1/m_b$.
When comparing the QCD result with the HQET result, we have no means
of disentangling the corrections to $f_B$ and $M_B$ from those to $B_i$,
so it is not obvious whether numerical discrepancies are 
due to corrections to the bag parameters, form factors, 
meson masses, the OPE, or even our
choices of sum rule parameters.  The conclusion to make
is that the leading-order expansion and the 
full results are  consistent with one another in 
all cases, within the uncertainties of the analysis.  

This said, further investigation of the HQET series for 
the OPE spectral densities for $R_2$ reveals some 
interesting features. As an example, we take the piece 
of the spectral density for $R_2$
multiplying $\langle GG \rangle$ as calculated with an axial-vector
current, and consider some higher-order terms in 
the $E_0/m_b$ expansion of the integrated spectral
density. Using the notation $x_i=E_i/m_b$, integrating over 
the square $0<x_i<x_0$, and normalizing to the
leading-order term in the $x_0=E_0/m_b$ expansion, we have
\be
\frac{4\int dx_1 dx_2 \rho^{\rm GG}
(x_1,x_2)}{ 4\int dx_1 dx_2 (-9/2 x_1 -9/2 x_2)}
=1-\frac{154}{27}x_0+\frac{560}{27}x_0^2+\dots =1.0-1.2+0.9+\dots
\ee
To derive the numbers we used $x_0=E_0/m_b\approx 0.2$ for $E_0=1~\GeV$.
The second and third terms are as large as the first, and the corrections
do not fall below 10\% until the sixth term, so 
the ``HQET'' expansion is not well behaved.  We put
HQET in quotes, because the expansion is just the diagrammatic one, not 
a rigorous one in terms of operators.
It would be interesting to see whether this poor convergence persists
even with a more careful treatment of the subleading corrections.
If so, this would have important implications for lattice QCD results,
where corrections to the HQET limit are not easy to control.

In quoting our final results, we used only those obtained from the
Borel sum rules.  However, one can work with either finite 
energy or Borel sum rules. Finite energy sum rules can be 
obtained from Borel sum rules 
in the limit $M^2\to\infty$ and are therefore more sensitive 
to the model of the continuum. We used both and saw little difference.
Our sum rule analysis is by no means unique.  
For instance, one can change the duality integrals by modifying 
each side of the sum rule in the same way 
(for instance by dividing both sides by $(s_1 s_2)$).
This definitely changes the shape of the curves and can provide 
better stability. However, our main point is that 
$\Delta B$ is so small that we 
need not be too sophisticated with the sum 
rules analysis. 
The splitting into factorized and non-factorized 
parts is powerful and useful precisely because 
the absolute value of $\Delta B$ turns out to be small.
Even with very conservative error estimates the results
are numerically informative, and our final results -- 
the range for the values of $\Delta B$ -- rather reliable. 

It is instructive to compare our approach to lattice QCD.  In
the lattice approach the parameter $B$ is computed as a whole,
since a splitting into factorizable and 
non-factorizable parts is not possible at the level of simulation. 
Then for the computation of the parameter 
$B$ (and not $\Delta B$ directly)
even good accuracy of the method (say, about 20\%, a 
typical accuracy in hadronic physics) 
gives a less precise statement about factorization than our
technique.

Our analysis was limited to leading order in perturbative 
corrections. A more accurate determination would require the computation 
of the next-to-leading order perturbative
contributions.  These involve three-loop diagrams and this 
is a non-trivial task.  Results are nonetheless available
for the operator $Q$ \cite{bbmixPT}, where it was 
shown that these corrections amount to  about 10\%.
For the other operators, we can say only that 
the corrections are parametrically on the order of
$\alpha_s/\pi$ and are also expected to be around $10\%$.
Thus, a qualitative prediction of the sum-rule 
analysis is that deviations from factorization are suppressed 
either by scale ratios or by the strong-coupling constant 
and are therefore small. 

\newpage
\section{Conclusions}
\label{sec:Conclusion}
We used QCD sum rules to calculate the bag parameters for the leading and
next-to-leading order operators in the $1/m_b$ expansion of
the transition operator used to analyze $B_s-\bar B_s$ mixing.
We found that the violation of the factorization approximation for
the matrix elements of both the leading and subleading operators due to 
non-perturbative vacuum condensate contributions is well under control and small. 
Our final results for the parameters $\Delta B_i=B_i-1$ are
\begin{eqnarray}
\Delta B |_Q & = & -0.6\pm 0.5 \% \nonumber \\
 \Delta B |_{Q_S} &=& -0.5 \pm 0.4 \% \nonumber \\
 \Delta B |_{R_2} &= &0.3 \pm 0.3 \% \nonumber \\
 \Delta B |_{R_3} & = &0.3 \pm 0.2 \% \nonumber 
\end{eqnarray}
We believe that our very conservative error estimates 
make our final quantitative results (the range for the values of $\Delta B$) 
rather reliable. 

Our result that the non-perturbative contributions to the $\Delta B_i$ are
extremely small means that the non-factorizable
contributions to the matrix elements are most likely dominated by calculable
perturbative effects.  We did not attempt to include the 
next-to-leading order perturbative contributions in our analysis.  
Our naive expectation, based on the existing
calculations for the operator $Q$, is that these corrections can contribute
an additional $\pm 10\%$.
To clarify this point would require 
the evaluation of the set of three-loop diagrams appearing in 
the perturbative analysis.

\vskip 1cm
{\bf Acknowledgements}
AAP thanks the Particle Theory Group of Siegen University 
where this work was
done during his stay as a Mercator Guest Professor 
(Contract DFG SI 349/10-1). BDP
acknowledges the support of the SFB/TR09 ``Computational Particle Physics''.
This work was supported by the German National Science Foundation (DFG)
under contract MA 1187/10-1 
and by the German Ministry of Research BMBF under contract 05HT6PSA.

\section{Appendix}
\label{sec:Appendix}
Here we compile the results for the condensate contributions to the 
OPE spectral density in each of the eight cases.
For the axial-vector current we single out the
scalar amplitude multiplying the structure tensor 
structure $p_1^\mu p_2^\nu$.
For the pseudoscalar interpolating current there is only one
amplitude as the correlation function is a scalar.

\subsection{Spectral densities for $Q$}
For the AV interpolating current we have
\ba
\Delta\rho_{\rm AV}&=&
\frac{1}{48\pi^2}\langle \frac{\alpha_s}{\pi}GG\rangle
\frac{1}{s_1  s_2}
(p_1 p_2) 2 z_1 z_2 (-3+z_1+z_2-2 z_1 z_2) \\
&\approx&
\frac{1}{48\pi^2}\frac{1}{4\pi^2}\frac{1}{m_b^2}
\bigg[ -6\langle g_s^2G^2\rangle \bigg]
\ea
where $z=m_b^2/s$ and we omit the factor $\theta(s_i-m_b^2)$ setting
the lower limits of integration for the $s_i$.  The 
factor $(p_1 p_2)= s_1/2+s_2/2$ for $q^2 = 0$. To take the heavy-quark 
limit in the second line we used $s_i=m_b^2(1+x_i)^2$ and 
expanded to lowest order in $x_i=E_i/m_b$. 

For the PS interpolating current:
\ba
\Delta\rho_{\rm PS}&=&\frac{1}{48\pi^2}\langle \frac{\alpha_s}{\pi}GG\rangle
\frac{1}{s_1  s_2}
(p_1 p_2)m_b^2 3 (-2+z_1+z_2- z_1 z_2)\\
&+&(p_1 p_2) \frac{1}{16\pi^2}\langle \bar s G s\rangle
m_b
\left(\frac{1}{s_1} 2z_1\delta(s_2-m_b^2)
+\frac{1}{s_2} 2z_2\delta(s_1-m_b^2)
\right)\\
&\approx&
\frac{1}{48\pi^2}\frac{1}{4\pi^2}\bigg[
-3 \langle g_s^2 G^2\rangle+12 m_b \pi^2 \langle \bar s G s\rangle
\left(2\delta(s_2-m_b^2)+
2\delta(s_1-m_b^2)\right)\bigg]
\ea

\subsection{Spectral densities for $Q_S$}

AV interpolating current:
\ba
\Delta\rho_{\rm AV} &=&\frac{1}{48\pi^2}\langle \frac{\alpha_s}{\pi}GG\rangle
\frac{m_b^2}{s_1 s_2}  \frac12 (6-3(z_1+z_2)+z_1 z_2)  \nonumber \\
&+&\frac{1}{16\pi^2}\langle \bar sG s\rangle
m_b \left(\frac{1}{s_1}(-2+z_1)\delta(s_2-m_b^2)
+\frac{1}{s_2}(-2+z_2)\delta(s_1-m_b^2)
\right) 
\\
&\approx & \frac{1}{48\pi^2}\frac{1}{4\pi^2}\frac{1}{m_b^2}
\bigg[\frac{1}{2} \langle g_s^2 GG\rangle
-12 m_b \pi^2 \langle \bar s G s\rangle
\left(\delta(s_2-m_b^2)+\delta(s_1-m_b^2)\right)\bigg]
\ea
PS interpolating current:
\ba
\Delta \rho_{\rm PS} &=& \frac{1}{48\pi^2}\langle \frac{\alpha_s}{\pi}GG\rangle
\frac{1}{s_1  s_2}
\left(
\frac{s_1s_2}{2}(6-3z_1-3z_2+z_1 z_2)
+(p_1 p_2)^2 z_1z_2 \right) \nonumber \\
&&+\frac{1}{16\pi^2}\langle \bar sG s\rangle
 m_b \left((-2+z_1)\delta(s_2-m_b^2)
+(-2+z_2)\delta(s_1-m_b^2)
\right)\\
&\approx & \frac{1}{48\pi^2}\frac{1}{4\pi^2}
\bigg[\frac{3}{2}\langle g_s^2 GG\rangle 
-12 m_b \pi^2 \langle \bar s G s\rangle
\left(\delta(s_2-m_b^2)+\delta(s_1-m_b^2)\right)\bigg]
\ea

\subsection{Spectral densities for $R_2$}

AV interpolating current:

\ba
&&\Delta \rho_{\rm AV} = \frac{1}{48\pi^2} 
\langle\frac{\alpha_s}{\pi}G G\rangle
\frac{1}{m_b^2}\bigg[\frac{1}{12}
(-4z_1^3z_2^3 + 12 z_1^3z_2^2 
- 4z_1^3z_2 - 9z_1^2z_2^2 -3z_1^2z_2 \nonumber\\
&&+ 9z_1z_2  
- 2z_1 + 1) \nn \\
&&+\frac{(p_1 p_2)^2}{s_1 s_2} z_1^2 z_2(2z_1z_2^2 - 
4z_1z_2 + z_1 + 4z_2 - 3)\bigg]+(z_1 \leftrightarrow z_2)\nonumber\\
&&+ \frac{1}{16\pi^2}\langle \bar sG s\rangle 
 \frac{1}{m_b}(\frac29 z_1^3 + \frac16 z_1^2 - \frac12 z_1+\frac19)
\delta(s_2-m_b^2)
+(z_1\leftrightarrow z_2) \\
&&\approx  
\frac{1}{48\pi^2} \frac{1}{4\pi^2}\frac{1}{m_b^2}
\bigg[-\frac{9}{2}x_1\langle g_s^2 G G\rangle 
- 12 x_1 m_b \pi^2 \langle \bar s G s\rangle \delta(s_2-m_b^2) + 
(x_1\leftrightarrow x_2)\bigg]
\ea
Note that the $1/m_b$ suppression of $R_2$ compared to $Q$ and $Q_S$
is manifest only after the HQET expansion.  The
coefficient of the $\langle g_s^2 G G\rangle$ term is large 
and there is a relative sign of $\langle G G\rangle$ and 
$\langle \bar s G s\rangle$ terms compared to all other cases.
This is a unique feature.

PS interpolating current:
\ba
&&\Delta \rho_{\rm PS} = \frac{1}{48\pi^2} 
\langle \frac{\alpha_s}{\pi}G G\rangle
\bigg[\frac{3}{8} ( - z_1^2z_2^2 + 3z_1^2z_2 
- z_1^2  - 2z_1z_2 -z_1 + 2) \nn \\
&&+\frac{3}{2} \frac{ (p_1 p_2)^2}{s_1 s_2} z_1 
(z_1 z_2^2 - 3z_1 z_2 + z_1  + 3z_2 - 2)\bigg]+(z_1\leftrightarrow  z_2) 
\nn \\
&&+ \frac{1}{16\pi^2}\langle \bar sG s\rangle 
 \frac{1}{m_b} \bigg[\left(\frac{m_b^2}{4} ( z_1^2 - 1) 
+ \frac{(p_1 p_2)^2}{m_b^2}  z_1^2 ( - z_1 + 1)\right)\delta(s_2-m_b^2)
\bigg] +(z_1\leftrightarrow z_2) \\
& &\approx  \frac{1}{48\pi^2}\frac{1}{4\pi^2}
\bigg[-\frac{3}{2}x_1 \langle g_s^2G G\rangle
+12 x_1 m_b\pi^2 \langle \bar sG s\rangle \delta(s_2-m_b^2)
+(x_1\leftrightarrow x_2)\bigg]
\ea
\subsection{Spectral densities for $R_3$}
AV interpolating current:

\ba
&&\Delta \rho_{\rm AV} 
=  \frac{1}{48\pi^2} \langle \frac{\alpha_s}{\pi}G G\rangle
\frac{(p_1 p_2) }{4 s_1 s_2}\bigg[
z_1( - z_1z_2^2 + 7z_1z_2 - 3z_1  - 9z_2 + 6)
\bigg]+(z_1\leftrightarrow z_2)\nonumber \\ 
&&+ \frac{1}{16\pi^2}\langle \bar sG s\rangle 
 \frac{(p_1 p_2)}{2 m_b s_1}(- z_1^2 +3 z_1- 2)\delta(s_2-m^2)
+(z_1\leftrightarrow z_2)\\
&&\approx  \frac{1}{48\pi^2} \frac{1}{4\pi^2}\frac{1}{m_b^2}
\bigg[\frac{1}{2} x_1 \langle g_s^2 G G\rangle 
- 12 x_1 m_b\pi^2 \langle \bar sG s\rangle \delta(s_2-m_b^2)
+(x_1\leftrightarrow x_2)\bigg]
\ea
PS interpolating current:
\ba
&&\Delta \rho_{\rm PS}= 
  \frac{1}{48\pi^2} \langle \frac{\alpha_s}{\pi}G G\rangle
 \frac{1}{m_b^2}\bigg[ \frac{(p_1 p_2)}{16}
(-z_1^2z_2^2 + 10z_1^2z_2 - 6z_1z_2  - 12z_1 + 9) \nn\\
&&+ \frac{(p_1 p_2)^3}{2s_1s_2} z_1^2z_2(-z_2 + 1)
\bigg] + (z_1\leftrightarrow z_2)\nn \\
 &&+ \frac{1}{16\pi^2}\langle \bar sG s\rangle 
\frac{(p_1 p_2)}{4 m_b} (-z_1^2 +3 z_1- 2)\delta(s_2-m_b^2)
+(z_1\leftrightarrow z_2)\\
&&\approx  \frac{1}{48\pi^2} \frac{1}{4\pi^2}
\bigg[ \frac{3}{4}x_1 \langle g_s^2 G G\rangle
-6 x_1 m_b \pi^2\langle \bar sG s\rangle \delta(s_2-m_b^2)
\bigg]+ (x_1 \leftrightarrow x_2)
\ea

\subsection{Four-quark condensates in the OPE}
In this subsection we discuss the treatment of the 4-quark condensates,
and show that their non-factorizable contributions to the current 
correlator vanish upon applying the vacuum saturation approximation.
The Fourier-transformed three-point correlators involving 
$Q$ and $Q_S$ are given by
\be
\label{input:q4}
T(p_1,p_2) = 2 \langle 0|(\bar s \Gamma_J (m_b+\pslash_1)\Gamma_O s)\,
(\bar s \Gamma_J (m_b-\pslash_2)\Gamma_O s)|0\rangle
\left(\frac{1}{(m_b^2-p_1^2)(m_b^2-p_2^2)}\right) ,
\ee
where $\Gamma_O= \gamma^\mu(1-\gamma_5)$ for $Q$ and $(1-\gamma_5)$
for $Q_S$, and $\Gamma_J=\gamma^\nu \gamma_5$ for the axial-vector
interpolating current, and $i\gamma_5$ for the pseudoscalar one.
In contrast to the case for the gluon and quark gluon-condensates, 
explicit expressions for the factorizable contributions to the 
three-point functions are needed.  They read 
\be
T^{Q}_{fac}=2\left(1+\frac{1}{N_c}\right)\Pi(p_1)\Pi(-p_2)
\ee
for $Q$ and 
\be
T^{Q_S}_{fac}=2\left(1-\frac{1}{2N_c}\right)\Pi(p_1)\Pi(-p_2)
\ee
for $Q_S$.  The functions $\Pi(p_i)$ for an operator with 
Dirac structure $\Gamma_O$ and an interpolating current
structure $\Gamma_J$ read
\be
\Pi(p)=i\int d^4x e^{ipx}
\langle 0|T \left\{\bar s \Gamma_J b(x) \bar b \Gamma_O s(0)\right\}|0\rangle
=\langle 0| \bar s \Gamma_J (m_b+\pslash)\Gamma_O s|0 \rangle 
\frac{1}{(m_b^2-p^2)} .
\ee
We now isolate the part of the 
current correlator $\Delta T$ which contributes to a
non-zero $\Delta B$ by using the definition (\ref{eq:deltaTdef}), which
gives
\be
\Delta T^i(p_1,p_2)=T^i(p_1,p_2)-
2\left(1+\frac{1}{N_c}\right)\Pi(p_1)\Pi(-p_2) .
\ee
The full correlator $T$ contains four-quark matrix elements of 
the form $\langle \bar s \Gamma_1 s \bar s \Gamma_2 s\rangle$.  Evaluating
these matrix elements using (\ref{eq:4quarkVA}), we find that 
$\Delta T$ vanishes for all cases.

\end{document}